\documentclass[sigconf]{acmart}

\makeatletter
\renewcommand\paragraph{\@startsection{paragraph}{4}{\z@}%
	{1.5ex plus .2ex minus .3ex}%
	{-0.5em}%
	{\normalsize\bf}}
\makeatother

\newcommand{\revtk}[1]{{\color{black}#1}}
\newcommand{\revtb}[1]{{\color{black}#1}}
\newcommand{\revtr}[1]{{\color{black}#1}}

\usepackage{enumitem,subcaption}
\usepackage{amsmath}
\usepackage{graphicx,algorithm,algcompatible,textcomp,tabularx}
\usepackage{hyperref}

\AtBeginDocument{%
	}

\setcopyright{acmcopyright}
\copyrightyear{2023}
\acmYear{2023}
\acmDOI{XXXXXXX.XXXXXXX}

\acmConference[Conference acronym 'XX]{Make sure to enter the correct
	conference title from your rights confirmation emai}{June 03--05,
	2018}{Woodstock, NY}
\acmPrice{15.00}
\acmISBN{978-1-4503-XXXX-X/18/06}

\begin{document}
	
	\title{VALERIAN: Invariant Feature Learning for 
	\revtb{IMU} Sensor-based Human Activity Recognition in the Wild}
	\author{Yujiao Hao}
	\affiliation{%
			\institution{McMaster University}
			\city{Hamilton}
			\state{Ontario}
			\country{Canada}
		}
	\email{haoy21@mcmaster.ca}
	
\author{Boyu Wang}
\affiliation{%
		\institution{Western Ontario University}
		\city{London}
		\state{Ontario}
		\country{Canada}
	}
\email{bwang@csd.uwo.ca}
	
\author{Rong Zheng}
\affiliation{%
		\institution{McMaster University}
		\city{Hamilton}
		\state{Ontario}
		\country{Canada}
	}
\email{rzheng@mcmaster.ca}
%
	
	\begin{abstract}
Deep neural network models for IMU sensor-based human activity recognition (HAR) \revtk{that are} trained from controlled, well-curated datasets suffer from poor generalizability in practical deployments. However, data collected from naturalistic settings often contains significant label noise. In this work, we examine two in-the-wild HAR datasets and DivideMix, a state-of-the-art learning with noise labels (LNL) method to understand the extent and impacts of noisy labels in training data. Our empirical analysis reveals that the substantial domain gaps among diverse subjects cause LNL methods to violate a key underlying assumption, namely, neural networks tend to fit simpler (and thus clean) data in early training epochs. Motivated by the insights, we design VALERIAN, an invariant feature learning method for in-the-wild wearable sensor-based HAR. 
By training a multi-task model with separate task-specific layers for each subject, VALERIAN allows noisy labels to be dealt with individually while benefiting from shared feature representation across subjects. We evaluated VALERIAN on four datasets, two collected in a controlled environment and two in the wild. Experimental results show that VALERIAN significantly outperforms baseline approaches. VALERIAN can correct 75\% -- 93\% of label errors in the source domains. When only 10-second clean labeled data per class is available from a new target subject, even with 40\% label noise in training data, it achieves $\sim 83\%$ test accuracy.  

	 Code is available at: 
	 \href{URL} {\revtk{https://github.com/YujiaoHao/VALERIAN.git}}
	\end{abstract}

	

	\maketitle
	
\section{Introduction}
Inertial measurement unit (IMU) sensor-based human activity recognition (HAR) has gained a lot of interest recently due to its pervasiveness in smartphones and smartwatch devices~\cite{yao2017deepsense,ordonez2016deepconvlstm,wang2019deep,nweke2018deep}. With the increasing adoption of deep neural network models in HAR tasks, there is a need to acquire a large amount of well-curated and labeled sensory data to train such models. Unfortunately, the majority of public HAR datasets are from controlled settings where subjects are asked to perform prescribed activities in lab environments. They typically contain a small collection of subjects and activity types over \revtk{a limited period} of time. For example, PAMAP2~\cite{reiss2012introducing}, a popular dataset for HAR, only includes eight subjects with 59.67 minutes of measurements per subject. Furthermore, data collected from controlled settings often have very different characteristics from those of freestyle motions in naturalistic environments~\cite{vaizman2017recognizing}.

Collecting IMU sensor data in the wild faces its own set of challenges. Arguably, the biggest difficulty is to label such data accurately~\cite{woodward2020labelsens}. Recalls from one's memory are known to be notoriously unreliable~\cite{ramirez2013creating}. Labeling wearable data by observing signal patterns requires extensive domain knowledge and experience since sensor readings are impacted by not only activity types but also subject characteristics, on-body positions and sensor orientations. A mainstream method to label such data is to resort to another human-interpretable modality such as visual or audio recordings and determine the labels manually post hoc. Unfortunately, labels obtained this way are still error-prone due to mis-synchronization across different modalities, human errors or missing data (e.g, occlusion in vision data). As the first contribution of the work, {\it we examine two datasets collected in naturalistic settings to understand the extent and characteristics of noisy labels.}

Learning with noisy labels (LNL) has long been investigated in the machine learning community with many effective methods being proposed for computer vision tasks. \revtb{Due to the lack of reliable ground truth in real-world noisy data, studies are mainly conducted by adding artificial noise to clean labeled datasets}~\cite{song2022learning}. Through an empirical study, we find that DivideMix, one state-of-the-art LNL method fails to achieve good accuracy and sometimes cannot converge at all. In-depth analysis reveals that the root cause is the violation of a key underlying assumption in \revtb{LNL} methods, i.e., models fit simpler (and thus clean) data in early training epochs. With substantial subject diversity, it is difficult to distinguish wrongly labeled data from correct ones from a different subject whose data follows a different distribution (also known as {\it domain gaps}). Therefore, the second contribution of the work is {\it to unravel the interplay between subject domain gaps and LNL for HAR tasks}. 

The insights from the empirical study motivate our third contribution, namely, the design of VALERIAN, an inVariant feAture LEarning foR In-the-wild domain AdaptatioN method of \revtb{IMU}-based HAR. Its core component is a one-step domain invariant feature learner that tackles label noises and learns the shared feature representation among multiple subjects \revtb{simultaneously}. VALERIAN uses self-supervised pretraining to learn robust features that are independent of label quality. 
The pretrained parameters are used to initialize the shared feature encoder of a multi-task learning model, where \revtb{each noisy labeled subject in the training set is considered as a separate task}. The network consists of shared feature encoder and subject-dependent task-specific layers that are trained iteratively \revtb{with noisy labeled data}. To combat noisy labels, early-learning regularization (ELR)~\cite{liu2020early} is adopted by introducing a loss term reflecting the temporal ensemble of past predictions. VALERIAN can be applied in two ways: 1) {\it label correction}, i.e., to clean the labels of noisy labeled datasets so that accurate HAR models can be developed, and 2) {\it domain adaption}, i.e., to adapt the trained model to an unseen subject. Specifically, VALERIAN can predict activity labels of each subject in the training set using a respective task-specific layer. 
\revtk{To achieve higher accuracy, we assume a small number of correctly labeled data is available from a new subject. The data is used to update a task-specific layer to allow fast adaption of the trained model to the subject.}

We evaluate the performance of VALERIAN using two controlled datasets with different levels and distributions of labeling noises, and two in-the-wild datasets. \revtk{Noises are introduced to investigate the impact of the amount of label noise on model performance. }VALERIAN consistently outperforms baseline approaches across all settings. \revtb{In label correction, VALERIAN can correct up to 93\% wrongly labeled samples. In domain adaptation, }even with 40\% label noise in training data, it achieves \revtk{an} $\sim83$\% test accuracy with only 10 seconds of correctly labeled data per class. A similar evaluation on a true in-the-wild dataset with noisy labels achieves an over 20\% improvement in the F1-Score compared to baseline methods. 

The rest of the paper is organized as follows. Section \ref{sec:motivation} describes the motivation of this work. In Section \ref{sc:method}, we introduce the VALERIAN method and the details of each component. In Section \ref{sc:evaluation}, we present the implementation details and performance evaluation of VALERIAN. Section \ref{sc:RelatedWork} describes the related work and how they differ from ours. Finally, we conclude the paper in Section \ref{sc:Conclusion} with a summary and \revtb{ a discussion of future research directions}.

\section{Motivation}
\label{sec:motivation}
To understand the characteristics of in-the-wild HAR datasets and to gain insights into why mainstream LNL methods tend to fail on such tasks, we inspect two datasets and the behavior of a state-of-the-art (SOTA) LNL algorithm in this section. 

\subsection{Characteristics of in-the-wild HAR datasets}
In this work, a HAR dataset is considered to be in the wild (or collected in naturalistic settings) if the activities of subjects are not precisely scripted. As a result, experimenters do not know exactly what activities shall be performed at what time. The ExtraSensory dataset is one such example~\cite{vaizman2017extrasensory}, where IMU data were collected from users' smartphone devices as they went about their daily activities. Activity labels were initially self-reported. Further curation was done by researchers who utilized information from other sensing modalities to automatically correct some labels. 
A detailed description of the curation procedure in ExtraSensory can be found in \cite{vaizman2017recognizing}. As another example, the Realworld dataset \cite{sztyler2019sensor} contains data collected from fifteen subjects performing activities such as climbing stairs down and up, jumping, lying, standing, sitting, running/jogging and walking. Although in most cases, subjects were asked to perform a certain activity, during \revtb{climbing up/downstairs} outside trials, the variations of terrains are not controlled by the experimenters and thus un-prescribed activities may occur. 

\begin{figure}[t!]
	\centering
	\includegraphics[width=0.58\linewidth]{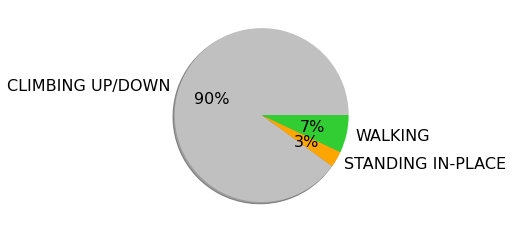}
	\includegraphics[width=0.4\linewidth]{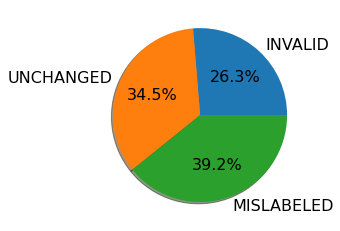}
	\caption{The statistics of two in-the-wild IMU-based HAR datasets. Left: Realworld, Right: ExtraSensory. A noticeable portion of the data labels in both datasets are noisy.}
	\label{fg:noisy_stats}
\end{figure}

\begin{figure}[t!]
	\centering
	\includegraphics[width=0.49\linewidth]{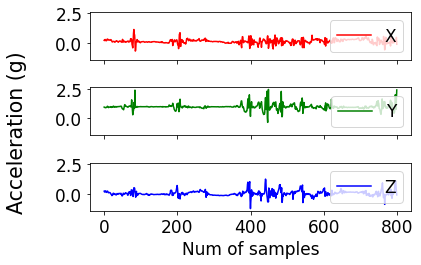}
	\includegraphics[width=0.49\linewidth]{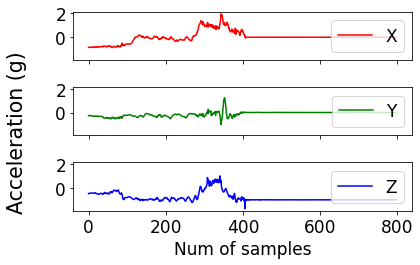}
	\caption{Accelerometer data in ExtraSensory with curated labels. Left: standing (subject id: FDAA70A1-42A3-4E3F-9AE3-3FDA412E03BF,
		row id: 4339), Right: walking (subject id: 2C32C23E-E30C-498A-8DD2-0EFB9150A02E, row id: 5454).}
	\label{fg:extrasensory}
\end{figure}

Fig.~\ref{fg:noisy_stats} illustrates the percentage of clean and mislabeled data in both datasets. For RealWorld, we inspect the video recording of climbing up and climbing down trials, note down the start and end times, and the type of activities. We find that there are periods \revtk{when} the subjects actually walk on flat ground (7\% of the time) or stand still (3\% of the time), which were mislabeled as climbing up or down in the dataset. 
For ExtraSensory, when comparing the self-reported and curated labels, we find that 34.5\% are unchanged, 39.2\% are corrected in the curation process and 26.3\% are marked as invalid since the phones were not \revtb{with the subjects during data collection}. Moreover, upon closer inspection of curated data in ExtraSensory, we find the data labels are still noisy. For example, in Fig.~\ref{fg:extrasensory}, the left plot corresponds to accelerometer measurements labeled as standing while the right one is labeled as walking. However, one can easily observe the ``signature" periodical pattern associated with walk cycles in the left plot \revtb{rather than} in the right plot -- an indication of mislabeling even after auto-curation. 

From Fig.~\ref{fg:noisy_stats}, we conclude ExtraSensory is much noisier than RealWorld since the former is crowdsourced data. What also distinguishes the two datasets is the distribution of \revtb{label noises}. Specifically, for RealWorld, most mislabeling happens in the climbing up/down trials when the ground labels are ``walk on a flat ground" or standing. In contrast, in ExtraSensory, mislabeling exists almost between any two activities. To characterize the distribution of noisy labels, a noise transition matrix $T$ is often used, where element $T_{ij}$ corresponds to the probability of mislabeling a data sample with ground truth label $i$ to label $j$~\cite{han2018coteach}. When mislabels occur equally likely for all classes other than the true class, the associate noise pattern is called {\it symmetric noise}. Otherwise, if there is a dominant off-diagonal element in each row in $T$,  the associate noise pattern is called {\it asymmetric noise}.

\begin{table}[]
\caption{The noise transition matrix of ExtraSensory, based on its curated labels. For walking and standing, only top-4 mislabeling sources are shown due to space limits.}
\label{tb:extra_confusion}
\resizebox{\linewidth}{!}{%
\begin{tabular}{l|lllll}
\hline
         & walking  & strolling & cleaning    & cooking       & eating   \\ \hline
walking  & 75.28\%  & 3.46\%    & 3.46\%      & 2.35\%        & 1.67\%   \\ \hline
         & running  & exercise  & go upstairs & go downstairs &          \\ \hline
running  & 79.92\%  & 19.66\%   & 0.21\%      & 0.21\%        &          \\ \hline
         & standing & cooking   & cleaning    & shower        & dressing \\ \hline
standing & 56.79\%  & 8.47\%    & 7.51\%      & 5.35\%        & 5.34\%   \\ \hline
         & at home  & at school & at work     & at party      & at gym         \\ \hline
at home  & 96.71\%  & 1.49\%    & 1.27\%      & 0.27\%        & 0.26\%         \\ \hline
\end{tabular}
}
\vspace{-2em}
\end{table}

Table~\ref{tb:extra_confusion} shows the noise transition matrix of data in three locomotion classes and one location class in ExtraSensory by comparing their curated labels (row headings) and the original ones (column headings). As ExtraSensory is a multi-label dataset with many classes, only top-5 mutually exclusive labels are included in the table. 
We observe that with the exception of ``running", noise transition probabilities of all classes are best modeled as symmetric noise.

\subsection{LNL can be Harmful to IMU-based HAR with noisy labels}
Learning with noisy labels has long attracted attention with many deep learning-based methods proposed recently that primarily target computer vision tasks. \revtb{According to \cite{song2022learning}, there are mainly four categories of LNL methods: robust architecture, robust regularization, robust loss design and sample selection. In this section, we use DivideMix~\cite{li2019dividemix}, a representative \revtb{sample selection} based method to illustrate the behavior and deficiency of LNL. In Section~\ref{sc:evaluation}, results from a robust regulation method are presented.}  

The basic idea of DivideMix is to first initialize a model with all training data for a few epochs (called {\it warm-up phase}). A Gaussian mixture with two modes is fitted to divide data samples based on their normalized losses into two partitions -- those with lower losses (higher confidence) are considered clean labeled samples, and those with high losses are treated as unlabeled data. Semi-supervised learning is then applied to the mixed data. Subsequently, co-refinement of labeled data and co-guessing of the labels of unlabeled data is performed by two neural networks working together iteratively, to reduce biases. 

To study the behavior of DivideMix for HAR, we \revtb{add artificial noise to clean labeled dataset. The USCHAD dataset~\cite{zhang2012usc} is selected as it contains carefully curated ground truth labels.} This dataset consists of accelerometer and gyroscope measurements collected from fourteen participants performing ten types of locomotions in a controlled environment (i.e., walking forward, walking left, walking right, walking upstairs, walking downstairs, running forward, jumping up, sitting, standing and sleeping). \revtb{Both symmetric and asymmetric noise patterns are considered, but due to space limits, only results from asymmetric noise are included. The transition matrix of asymmetric noise is defined by flipping pairs of the most confusing activities (see Fig. \ref{fg:asym_t} for details).}
We adopt the DeepConvLSTM model architecture proposed in \cite{ordonez2016deepconvlstm} \revtb{as feature extractor} for HAR tasks. The model contains 4 convolutional neural network (CNN) layers and 2 long short-term memory (LSTM) layers totalling $\sim$296k trainable parameters.

\begin{figure}[h!]
	\centering
	\begin{subfigure}{0.45\linewidth}
	\includegraphics[width=\textwidth]{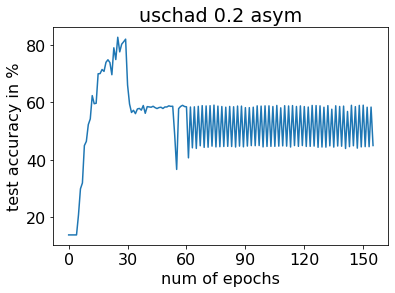}
	\caption{Test accuracy}
	\label{fig:dividemix_sym0.1_error}
	\end{subfigure}
	\begin{subfigure}{0.45\linewidth}
	\includegraphics[width=\textwidth]{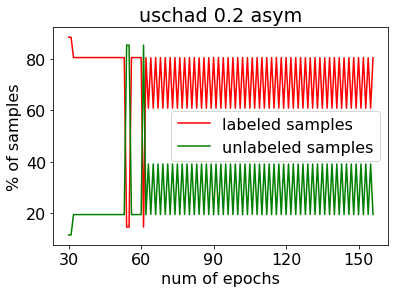}
    \caption{Division of clean and noisy labeled data}	
    \label{fig:dividemix_sym0.1_label}
	\end{subfigure}

	\caption{The performance of DivideMix on USCHAD in leave-one-subject-out experiments.}
	\label{fg:dividemix}
\end{figure}

Fig.~\ref{fg:dividemix} shows the behavior of DivideMix over training epochs in presence of 0.2 asymmetric labelling noise. In the experiments, 13 of 14 subjects are included in the training data and the remaining subject is used in testing. The warm-up phase ends at 30 epochs. 
As shown in Fig.~\ref{fig:dividemix_sym0.1_error}, test accuracy increases quickly during the warm-up phase indicating that the model can learn despite label noises. However, after the warm-up phase, the test accuracy drops drastically and fluctuates between 45\% and 60\% after 60 epochs. A closer look at the division between labeled and unlabeled data in the training set is in Fig.~\ref{fig:dividemix_sym0.1_label}. It reveals that despite only 20\% of the data samples being labeled incorrectly, DivideMix gradually converges to split the data approximately 81-19 or 61-39. As a result, some clean labeled data is classified as unlabeled and fail to contribute as much to the training process. 

\begin{figure}[h!]
	\centering
		\includegraphics[width=0.49\linewidth]{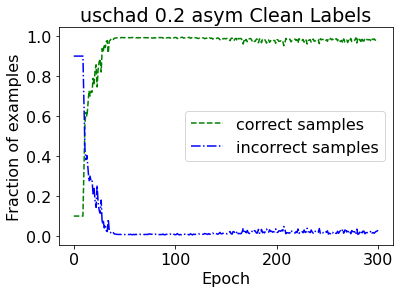}
		\includegraphics[width=0.49\linewidth]{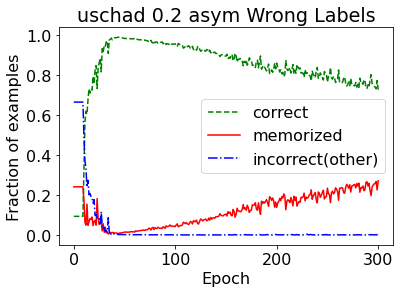}

\caption{Results of the DivideMix model on USCHAD with 0.2 asymmetric noise. Left: the fraction of clean labeled samples that are predicted correctly (green) and incorrectly (blue). Right: the fraction of samples with wrong labels that are predicted correctly (green), memorized (red), and incorrectly as neither the true nor the labeled class (blue).}
\label{fig:USCHAD_memorization}
\end{figure}

To shed the light on why DivideMix fails in these experiments, further analysis is in order. 
First, we inspect the effect of memorization. Deep neural network models are known to have the propensity for fitting training data including outliers or mislabeled data. However, it has been empirically demonstrated that such a memorization phenomenon tends to happen at a late stage of training~\cite{arpit2017mem,liu2020early}. In early training epochs, the model prioritizes learning simple patterns. 
To test if this hypothesis is true for HAR tasks, we show in Fig.~\ref{fig:USCHAD_memorization} the breakdown of training samples among five categories. Specifically, a data sample that is correctly labeled can be either correctly or wrongly predicted by the trained model up to the associated epoch. For a data sample that is wrongly labeled, three situations may arise: i) its prediction is the same as the ground truth label ({\it correct}), ii) its prediction is the same as the wrong label ({\it memorized}) or iii) otherwise, i.e., its prediction is neither the ground truth label nor the wrong label. From Fig.~\ref{fig:USCHAD_memorization}, even after a few epochs, memorization is non-negligible, especially in the case of asymmetric noise. When a noisy label is memorized, the model has high confidence in its {\it wrong} prediction. 

	
	\begin{figure}[h!]
	\centering
		\begin{subfigure}{0.32\linewidth}
		\includegraphics[width=\textwidth]{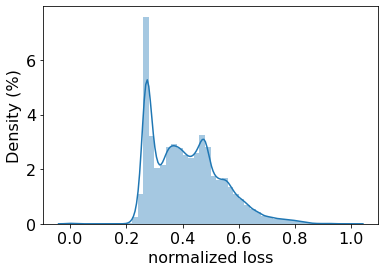}
		\caption{Distribution of normalized losses}
		\end{subfigure}
		\begin{subfigure}{0.32\linewidth}
		\includegraphics[width=\textwidth]{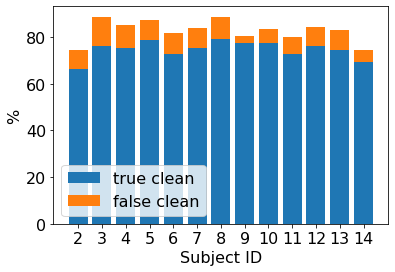}
		\caption{Partition of data predicted as clean}
		\end{subfigure}
		\begin{subfigure}{0.32\linewidth}
		\includegraphics[width=\textwidth]{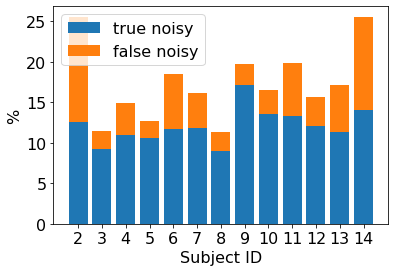}
		\caption{Partition of data predicted as noisy}
		\end{subfigure}
		\caption{Effects of subject diversity on early learning. Plots are generated on a model trained on Subject 2 -- 14 in USCHAD with 0.2 asymmetric noise and after 30 epochs of warm-up training in DivideMix.}
		\label{fg:root_cause2}
	\end{figure}

We believe the root cause of early memorization and the consequent failure of DivideMix in HAR tasks is due to the large variability across subjects when performing the same activity. Subject diversity is a well-recognized problem in IMU-based HAR~\cite{chen2021deep}. However, the problem is exacerbated when noisy labels are present. In Fig.~\ref{fg:root_cause2}, we show the normalized cross-entropy losses for Subject 2 -- 14 in the training data and the division of clean and noisy labels for each subject in DivideMix after a 30-epoch warm-up period. 
Clearly, the normalized losses (Fig.~\ref{fg:root_cause2}(a)) no longer follow a two-component GMM. Instead, they are better modelled by a mixture of three or more components. 
Inspecting the division of labelled and unlabeled data for each subject by DivideMix, we find that some \revtk{data presumed to be clean} is in fact noisy (\revtb{Fig. \ref{fg:root_cause2}(b)}) while a  portion of presumably noisy data is in fact clean for each subject (false noisy in \revtb{Fig.~\ref{fg:root_cause2}(c)}). Some subject (e.g., Subject 14) appears to be penalized with a higher percentage of clean data being mislabeled as unlabeled by DivideMix. More than 10\% of Subject 14's clean data is \revtk{misclassified} as \revtk{noisy} (due to high normalized losses). 
	\begin{figure}[h!]
	\centering
		\begin{subfigure}{0.32\linewidth}
		\includegraphics[width=\textwidth]{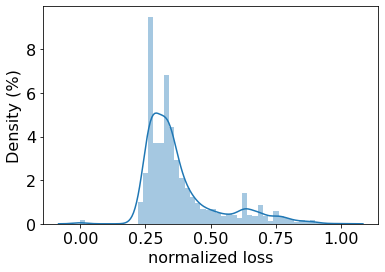}
		\caption{Distribution of normalized losses}
		\end{subfigure}
		\begin{subfigure}{0.32\linewidth}
		\includegraphics[width=\textwidth]{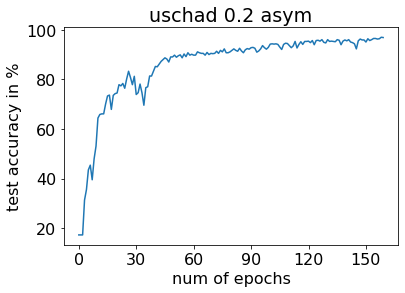}
		\caption{Test accuracy}
		\end{subfigure}
		\begin{subfigure}{0.32\linewidth}
		\includegraphics[width=\textwidth]{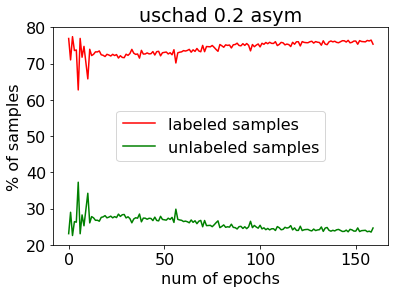}
		\caption{Division of clean and noisy labeled data}
		\end{subfigure}
		\caption{\revtk{Performance of DivideMix on a single subject (subject id: 2). (a) is generated after the warm-up phase, while (b) and (c) are generated on the full training process of a DivideMix model.  }
		}
		\label{fg:single}
	\end{figure}

Fig.~\ref{fg:single} shows the \revtk{case when training DivideMix on data from one subject with 0.2 asymmetric labeling noise. We train the model using data from 4 trials of the subject and test with the remaining trial.} To avoid overfitting, the network size of DeepConvLSTM is reduced \revtb{by retaining only two CNN layers and one LSTM layer with a total of 56k trainable parameters.} The distributions in Fig. \ref{fg:single} can indeed be modeled as 2-component GMM following the basic assumption of DivideMix and thus can be correctly handled by the method (results omitted for brevity).
Comparing the results \revtk{from Fig.~\ref{fg:single}} with Fig.~\ref{fg:root_cause2}(a) \revtk{and Fig.~\ref{fg:dividemix}}, it is \revtk{clear that DivideMix works \revtk{reasonably} well on data from  a single subject but failed in the case of multiple subjects. Therefore, it is} reasonable to conclude that the discrepancy is due to the domain gaps in multi-subject data. 

Though our analysis focuses on DivideMix, other categories of LNL methods such as \revtb{ELR~\cite{liu2020early} and CDR \cite{xia2020robust}} make the same assumptions that high-confidence labels in early training stages are more trustworthy. Unfortunately, as evident from the empirical analysis in this section, such assumptions no longer \revtk{hold} in presence of diverse subject data in HAR tasks. 
\section{Method}
\label{sc:method}
\begin{figure*}[h!]
\centering
\includegraphics[width=0.8\linewidth]{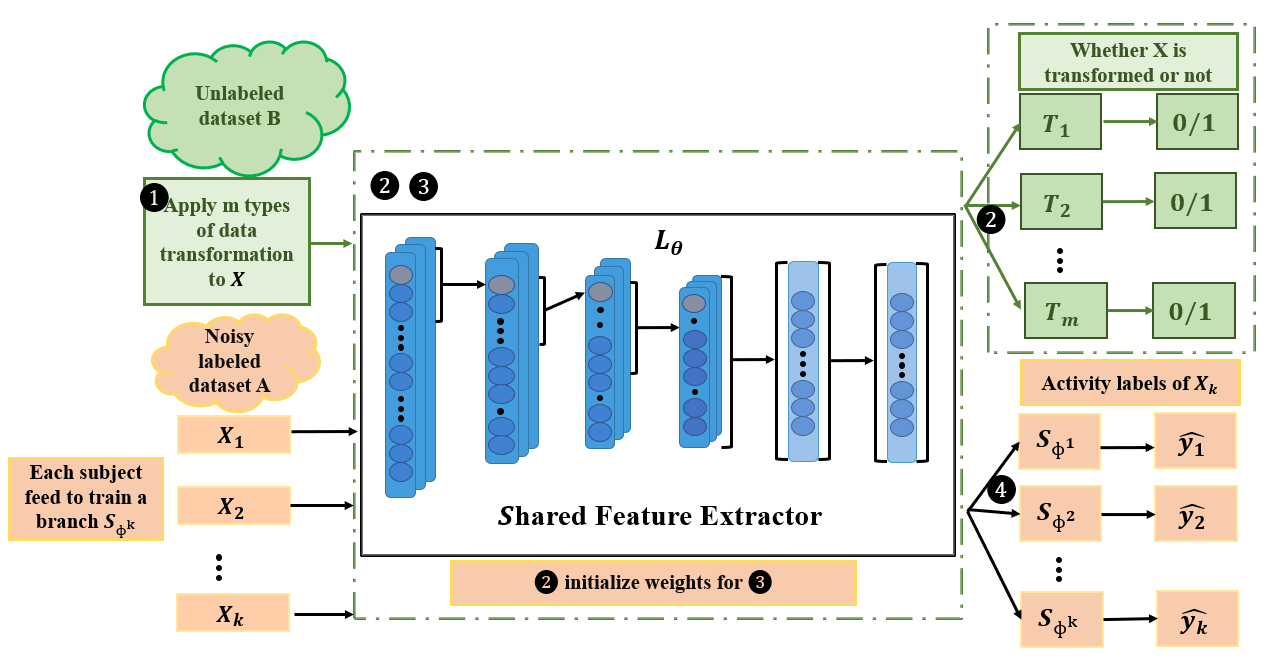}
\caption{\revtk{An overview of VALERIAN training procedure. Step 1: Data transformation. Step 2: self-supervised training. Step 3 and 4: alternating training.} 
}
\label{fg:overview}
\end{figure*}
Let the input and label spaces be $\mathcal{X}$ and $\mathcal{Y}$,  respectively. Due to high subject diversity in HAR tasks, each subject in the training set is treated as a separate source domain in the joint space $\mathcal{X}\times\mathcal{Y}$. 
In the rest of the paper, we use ``domain" and ``subject" interchangeably. Let $\mathcal{D}_{k} = \{(x^{k}_{n},\tilde{y}^{k}_{n})\}^{{N}_{k}}_{n=1}$, where $N_{k}$ is the number of data samples from subject $k$ and $\tilde{y}$ denotes  noisy labels.
The source domains are denoted by $\mathcal{D}_{s}=\{\mathcal{D}_{1}, \mathcal{D}_{2},..., \mathcal{D}_{K}\}$, where $K$ is the number of subjects. We further assume that a small collection of clean labeled samples can be obtained for an unseen subject $t$ denoted by $\mathcal{D}_{t}=\{(x_n,y_n)\}^{M}_{n=1}$. The goal of {\it HAR from data in-the-wild} is to learn a model from $\mathcal{D}_{s}$ that can either be easily adapted to a new target domain given $\mathcal{D}_{t}$, or be used to correct the wrong labels in $\mathcal{D}_{s}$.


Motivated by the observations from Section~\ref{sec:motivation}, we propose VALERIAN, a one-step method that handles noisy labels and distribution gaps across multiple source domains simultaneously. 
Our solution is based on two key insights: i) unsupervised learning that aims to learn representations invariant to instance-level variations is not affected by noisy labels; and ii) within each source domain, clean data tends to exhibit simpler patterns (than wrongly labelled data), which can be learned in early training epochs. Moreover, we assume that in absence of noisy labels, there exist domain-invariant features across subjects in HAR tasks. This assumption has been empirically verified in prior work~\cite{hao2021invariant}. 
\revtb{After model training, VALERIAN can be used in both cleaning noisy labels in the training data and enabling fast adaptation to a new subject from a small amount of clean labeled data (Fig.~\ref{fg:overview}).}

\subsection{Solution overview}
VALERIAN takes advantage of known techniques in machine learning but combines them in innovative ways. It has three key building blocks: i) self-supervised pre-training, ii) invariant feature learning from noisy labelled data, and iii) fast adaption to unseen subjects. 

Self-supervised pre-training takes unlabeled data and performs data augmentation to pre-train feature extractor that captures structures of underlying distributions. Invariant feature learning in VALERIAN has two objectives: 1) to learn shared feature representations across domains and 2) to combat the memorization effect introduced by noisy labels. To do so, we adopt a multi-task learning model for domain invariant feature learning which was first proposed in \cite{hao2021invariant}. The model consists of a shared feature extractor across multiple source domains and multiple task-specific output layers.  
To counter the effect of noisy labels, we introduce a regularization term similar to ELR in the loss function during training. 
 Finally, for a new subject with a small amount of clean data, fast adaption is performed on one of the task-specific layers only.
 
 Algorithm 1 summarizes the training procedure of VALERIAN. Next, we will provide the details of each building block. 

\begin{algorithm}[t]
	\caption{Invariant feature learning for in-the-wild domain adaptation}
	\begin{algorithmic}[1]
		\REQUIRE
		Source domains $\mathcal{D}_{s} = \{D_{k}\}_{k=1}^{K}$, learning rate $\gamma$, hyperparameters $\alpha,\beta, \lambda,\mu$
		\ENSURE
		VALERIAN model with parameter $\theta$ and $\phi$
		\State Initialize $\theta$ with self-supervised pretrain
		\State Random initialize $\phi=\{\phi^{1},\phi^{2},...,\phi^{K}\}$
		\State Initialize ensemble predictions $t \leftarrow 0_{[n \times C]}$
		\REPEAT 
		\State Sample tasks $T=\{T_{1},T_{2},...,T_{K}\}$ over  $\mathcal{D}_{s}$
		\State //Update $\phi^{k}$ with fixed $\theta$
		\FOR {$k$ is 1 to $K$}
		\State Freeze parameters of $\phi$ except $\phi^{k}$
		\hspace{\parindent} \FOR {each minibatch B in $T_k$}
		\FOR { $(x_i,\tilde{y}_i)$ in B}
		\State $p_i \leftarrow S_{\phi^k}(L_{\theta}(x_i))$
		\State $t_i \leftarrow \beta t_i+(1-\beta)p_i$
		\ENDFOR
		\ENDFOR 
		\State $\mathcal{L}oss \leftarrow \mathcal{L}_{CE}(T_k,\theta;\phi^k)+\mu|\phi^k|_1+
		 \frac{\lambda}{|B|}\sum \log(1- \langle p_i, t_i \rangle )$
		\State $\phi^{k}\leftarrow\phi^{k}-\gamma \nabla_{\phi^{k}}\mathcal{L}oss(T_{k},\theta;\phi^{k})$
		\ENDFOR
		\State //Update $\theta$ with fixed $\phi$
		\FOR {each minibatch B in $T$}
		\State $B'=Mixup(B,\alpha)$
		\FOR { $(x_i,\tilde{y}_i)$ in B'}
		\State  $p_i \leftarrow S_{\phi}(L_{\theta}(x_i))$
		\State $t_i \leftarrow \beta t_i+(1-\beta)p_i$
		\ENDFOR
		\ENDFOR
		\State $\mathcal{L}oss \leftarrow \mathcal{L}_{CE}(T,\phi;\theta)+\mu|\phi|_1+
		\frac{\lambda}{|B|}\sum \log(1- \langle p_i, t_i \rangle )$
		\State  $\theta\leftarrow\theta-\gamma \nabla_{\theta}\mathcal{L}oss(T,\phi;\theta)$  
		
		\UNTIL convergence
	\end{algorithmic}
\end{algorithm}

\subsection{Self-supervised pre-training}
\label{sc:self-supervised}
In \cite{jiang2020beyond}, the authors find that a ResNet pre-trained on ImageNet datasets appears to work consistently better than random initialized ones as a feature extraction network for LNL image classification tasks.
Inspired by this, we pre-train a feature extractor network by removing the labels in HAR datasets. It is thus natural to consider feature learners that require no label information, such as contrastive learning \cite{chen2020simclr} or self-supervised learning. Self-supervised learning is a machine learning method that learns semantic features from unlabeled data with customized tasks~\cite{doersch2017multi}. As there is no ground truth label, to \revtk{leverage} of this technique,
data augmentation \revtk{techniques} and auxiliary tasks need to be introduced. In  \cite{saeed2019multi}, Saeed {\it et al.} introduce \revtk{various data transformations} and train a multi-task model to classify the type of transformation applied. \revtk{The features extracted from the IMU data embed information regarding natural human motion while the transformed ones introduce different degrees of distortion. Trained  to classify the type of transformation applied, a neural feature extractor \revtk{learns to represent human motion more accurately} and obtains more meaningful discriminative features.} We adopt the same idea and apply the following transformations: 

\begin{enumerate}
\item {\it Noised}: it adds random Gaussian noise to the original data samples. 
\item {\it Scaled}: this transformation changes the magnitude of data samples within a sliding window by multiplying with a random scalar.
\item {\it Rotated}: this transformation mimics different sensor orientations by multiplying the original data with a rotation matrix of randomly generated axis-angle.
\item {\it Negated}: this transformation negates samples within a time window, resulting in a vertical or a horizontal flip of the original input signal.
\item {\it Reversed}: it reverses the data along the \revtk{time-axis}, resulting in a complete mirror image of the original input.
\item {\it Permuted}: sensor signals are randomly sliced and swapped within a data window.
\item {\it Time-Warped}: it mimics the change of motion frequency by locally stretching or warping a time series through a smooth distortion of time intervals.
\item {\it Channel-Shuffled}: it randomly shuffles sensor data in axial dimensions.
\end{enumerate} 
One or several of these transformations (called {\it pretext tasks}) are applied to each data window of each sensor separately (accelerometer and gyroscope). 
Each head of the multitask learning model corresponds to a binary classifier. By learning whether a certain type of transformation has been applied to the original data samples, the feature extractor portion of the network captures high-level semantic information that is invariant to these transformations and thus beneficial to downstream tasks. 

\subsection{Domain invariant feature learning}
\label{sc:BMTL}
Self-supervised learning alone is insufficient to handle domain gaps among subjects. Moreover, data labels are  necessary to fine tune model parameters for downstream tasks. To generalize well to unseen subjects, we utilize the invariant feature learning framework (IFLF) from \cite{hao2021invariant} but modify it to work with LNL. It consists of three components:
\paragraph*{\textbf{Alternating training}} An IFLF model is a multi-task model trained with tasks sampled from all source domains. Each subject has its individual task-specific layer $S_{\phi^k}$ but shares a common feature extractor network $L_\theta$. If the model is trained by simply iterating among tasks sampled from $\mathcal{D}_{1}$ to $\mathcal{D}_{K}$, catastrophic forgetting may occur\cite{kirkpatrick2017overcoming}, namely, a model forgets previously learned tasks, and can only work properly on newly learned tasks. To avoid catastrophic forgetting, the alternating training strategy is employed from \cite{kumar2012learning}, to update $L_{\theta}$ and $S_{\phi^{k}}$ separately. In each training epoch, we first freeze the parameters of the feature extractor network, and update the parameters of each task-specific layer with its respective data; then, we freeze the parameters of all task-specific layers and update the invariant feature extractor using all data from the previous step.
\paragraph*{\textbf{Feature extractor}} By the merit of multi-task learning, $L_{\theta}$ generalizes well across domains through the shared representations among related tasks \cite{ruder2017overview}. For HAR tasks, we use DeepConvLSTM \cite{ordonez2016deepconvlstm} as the backbone network. 
It includes four CNN layers and two LSTM layers. 

The objective function $\ell_L$ works on multiple source domains to learn a domain invariant feature representation that clusters the features by their labels. It is defined as follows:

\begin{equation}
	\label{eq:loss_l}
	\ell_L = \sum_{k=1}^{K}\mathcal{L}_{CE}(T_k,\phi^k;\theta),
\end{equation} 
where $\mathcal{L}_{CE}$ is the categorical cross-entropy loss function calculated on each $T_{k}$ with given $\theta$ and $\phi$, defined as
$\mathcal{L}_{CE}=-\sum_{i=1}^{C}  \tilde{y}_i log(p_i)$
on data from each task $k$. We call such a multi-task model \textit{basic multi-task learning model} (BMTL). To further boost the quality of extracted features, we use self-supervised pre-train as described in Section \ref{sc:self-supervised} to initialize the model parameter $\theta$.

\paragraph*{\textbf{Task-specific networks}} Generally, if the shared feature generalizes well across all source domains, it also works well on the target domain. $L_{\theta}$ needs to have sufficient capacity to explore the entire latent space $\mathcal{Z}$ and extract domain invariant features. In contrast, a task-specific network $S_{\phi}^{k}$ should be as simple as possible with fewer learnable parameters to allow fast adaptation with target domain data. In the implementation, a lightweight task-specific layer $S_{\phi^{k}}$ includes a fully connected layer with a softmax activation function. The task-specific objective function is defined as the sum of a categorical cross-entropy loss and an $\ell_1$-norm regularization term as follows,
\begin{equation}
	\label{eq:loss_s}
	\ell_{S^{k}} = \mathcal{L}_{CE}(T_k,\theta;\phi^k) + \mu |\phi^{k}|_{1}, k = 1, 2, \ldots, K, 
\end{equation}
where $\mu$ is a hyper-parameter to control the sparsity of $S_{\phi}^{k}$. The regularization term imposes sparsity on the task-specific layers and helps mitigate overfitting. 
%
\subsection{Learning with noisy labels}
With the multitask learning model introduced previously, we can get the best of both worlds: shared network parameters for feature extraction for all subjects and subject-dependent output layers. As a result, the underlying assumption of dominant LNL methods is that in early training epochs, each subject-dependent model tends to incur low losses (higher confidence) on clean data and large losses on mislabeled data are likely to hold. To handle noisy labels, in principle, we can incorporate any existing LNL method in the invariant feature learning framework. However, we find that DivideMix has high training costs due to its use of two networks in co-teaching and co-refinement. When combined with invariant feature learning, its complexity grows linearly with the number of source domains. Therefore, in VALERIAN, we use ELR to counter memorization effects by forcing model predictions to be close to their temporal ensemble. An ELR loss is defined as : 
\begin{equation}
	\label{eq:elr}
	\mathcal{L}_{elr}= \frac{1}{|B|}\sum_{i=1}^{|B|} \log\left(1- \langle p_i, t_i \rangle \right),
\end{equation}
where $p_i$ is the model output of input sample $x_i$, and $t_i=\beta t_i + (1-\beta)p_i$ is the temporal ensemble controlled by hsyper-parameter $\beta$. \eqref{eq:elr} maximizes the inner product of $p_i$ and $t_i$, and the logarithm in $\mathcal{L}_{elr}$ inverts the exponential function implicit in the softmax function in $p_i$. 

MixUP~\cite{zhang2017mixup} is a simple yet effective data augmentation technique in \revtk{improving} model generalization capabilities~\cite{xu2020adversarial}. In HAR tasks, we can mix up data samples from the same activity class but different subjects. To apply Mixup data augmentation, each data sample of a mini-batch is interpolated with another sample randomly chosen from a different source domain but belongs to the same class. Specifically, for a pair of samples $(x_1,\tilde{y}) \in \mathcal{D}_i$ and $(x_2,\tilde{y}) \in \mathcal{D}_j$, the mixed data sample $(x',\tilde{y})$ is computed by:
\begin{flalign}
	a \sim Beta(\alpha,\alpha),\\
	a' = max(a, 1-a),\\
	x'=a'x_1+(1-a')x_2
\end{flalign}
where $a$ is the MixUp factor sampled from a $Beta$ distribution controlled by hyper-parameter $\alpha$.
Finally, the total losses in \eqref{eq:loss_l} and \eqref{eq:loss_s} are updated as:
\begin{equation}
	\mathcal{L}oss_{L} = \ell_{L}+\mu|\phi|_1+\lambda \mathcal{L}_{elr},
\end{equation}
\begin{equation}
	\mathcal{L}oss_{S^k} = \ell_{S^k}+\lambda \mathcal{L}_{elr}, k = 1, 2, \ldots, K,
\end{equation}
where $\lambda$ is a hyper-parameter to control the importance of ELR. It is worth noting that the loss is calculated differently in the alternating training procedure as $L_\theta$ includes all source domains while $\phi^k$ only concerns the data of the $k$th subject. MixUp augmentation is only used in updating the feature extraction layers ($L_\theta$).

\subsection{Applications of VALERIAN}
\subsubsection{Label correction}
After learning the domain invariant features from a noisy training set, VALERIAN is capable of relabeling the training samples close to their ground truth activities. For a (noisy labeled) data sample $(X,\tilde{y})$ from subject $k$, the prediction $\hat{y}^k$ of task-specific layer $S_{\phi^k}$ is taken as the new label for the sample.
\subsubsection{Fast adaptation to new subjects} 
Since the network parameters in task-specific layers are already sparse, for a new subject, one can either initiate a new task-specific layer from scratch or randomly select a $S_{\phi^k}$ to update its trained parameters. A small amount of clean data is taken from $\mathcal{D}_t$ to train the task-specific layer.
\section{Performance Evaluation} 
\label{sc:evaluation}

\subsection{Datasets}
We consider four \revtk{publicly available datasets} to cover a wide variety of device types, data collection protocols, and activity classes \revtk{for recognition}. Because the evaluation of machine learning models requires the availability of clean ground truth labels, the first two datasets, USCHAD and WISDM~\cite{weiss2019smartphone} were collected under controlled laboratory environments. To simulate labelling errors, symmetric or asymmetric noise is injected into the labels with different noise transition matrices. WISDM contains a large number of subjects. Raw accelerometer and gyroscope data were collected from a smartphone in each participant's \revtk{pants} pocket at a rate of 20Hz. There are a total of 51 test subjects performing seven locomotion activities (i.e., walking, jogging, stairs, sitting, standing, kicking a soccer ball, playing tennis) for three minutes per trial to \revtk{achieve} equal class distribution. 

The third and fourth datasets, ExtraSensory and RealWorld, allow us to gauge VALERIAN's ability to handle real in-the-wild data. In ExtraSensory, crowdsourced mobile phone data are collected from 60 subjects during daily living activities. In the evaluation, we only consider six locomotion-related activities, namely, walking, running, cycling, sitting, standing and lying down. In the absence of ground truth labels, we take instead the curated \revtk{data} labels as ground truth. However, as discussed in Section~\ref{sec:motivation}, the curated data remains to be noisy. Moreover, ExtraSensory also suffers from severe class imbalance and missing class issues (only nine out of 60 subjects have data from all six classes in the dataset).


\subsection{Baseline methods}
Five baseline models have been implemented for comparison. 
\begin{itemize}
    \item {\it Single-task learning model (STL)}: STL is trained from scratch solely on the clean data from a target domain (a new subject). As the number of clean data increases, it is \revtk{expected STL's performance to improve} since there is no label noise.
    \item {\it Basic multi-task learning model (BMTL)}: Similar to VALERIAN, BMTL is a multi-task learning approach trained with noisy source domains and adapted to a target domain with a small number of clean labels. However, unlike VALERIAN, BMTL does not perform self-supervised pre-training and treats all training data \revtk{as if it were} clean. 
    \item {\it Subject-independent model with cross-entropy losses (SI)}: It pools all but test subjects' data to train a single subject-independent model and treats all training data as clean. 
    \item {\it Subject-independent model with ELR (SI-ELR)}: It is a subject-independent model trained by pooling all but test subjects' data together. Unlike SI, it utilizes ELR to combat noisy labels. Additionally, we denote by {\it SI-ELR-best}  the best-performing model (based on clean validation data) saved after the training epochs. Note, in practice, we cannot decide when to stop training to obtain SI-ELR-best with truly noisy data, and thus its results are presented for reference only.
    \item {\it Butterfly~\cite{liu2019butterfly}}: It is a joint LNL and domain adaptation method, which treats all but test subjects' data as a single source domain. It takes all unlabeled data samples from a target domain together with noisy labeled source domain data to train a model. Butterfly maintains four deep networks simultaneously, two for adaptations (i.e., noisy-to-clean, labeled-to-unlabeled, and source-to-target domains) and the remaining two for classification in the target domain.
\end{itemize}
STL, BMTL and VALERIAN are all \revtb{supervised domain adaptation} methods and utilize some data from the target domain. In contrast, SI and SI-ELR do not require any target domain data. 
Butterfly on the other hand includes unlabeled target domain data during training and thus \revtk{no transfer learning is done at inference time using labeled target domain data}. 

\subsection{Implementation and evaluation procedure}
\label{sc:experiment_setup}
\paragraph*{\textbf{Data preparation}} A standard IMU data pre-processing procedure is implemented for the experiments, including interpolation, low-pass filtering, normalization, and data segmentation. A Butterworth low-pass filter \cite{butterworth1930theory} with a cut-off frequency of 10Hz is employed to remove high-frequency noise from interpolated data. After low-pass filtering, we normalize the data and then segment it into sliding windows with a fixed length of 2 seconds with \revtk{an 80\% overlap} between adjacent windows. 
\paragraph*{\textbf{Implementation}}
The implementation of the feature extractor follows DeepConvLSTM in all models. It includes four layers of 1D CNN and two LSTM layers with 128 hidden units and a dropout rate of 0.25 to prevent over-fitting \cite{srivastava2014dropout}. The CNN layers have 64 channels with kernel size 5 and stride 1.     

For STL, the models are trained with a RMSProp optimizer \cite{bengio2015rmsprop} at a learning rate of $10^{-3}$ and a decay factor of $p=0.9$. The maximum iteration number is set to be 500. The SI models are trained with 200 epochs only, as the memorization effect will gradually degrade the model performance in latter training epochs. Butterfly and ELR are trained using hyper-parameters as specified in the original papers. VALERIAN utilizes DeepConvLSTM in $L_{\theta}$ while the number of $S_{\phi^{k}}$ branches is determined by the number of subjects in the training data. Each $S_{\phi^{k}}$ may have a different output shape depending on the number of classes in the dataset for the \revtk{corresponding} subject. VALERIAN is trained with an Adam \cite{kingma2014adam} optimizer at a learning rate of $10^{-4},\beta_{1}=0.9,\beta_{2}=0.999$, with hyper-parameters $\mu=0.4$, $\alpha=0.2$, $\beta =0.7$, and $\lambda=3$. The batch size is set to 64 and the number of training epochs is 300 without early stopping. The hyper-parameters and the  optimizer used in each model are consistent across all datasets.

\paragraph*{\textbf{Evaluation process}} 
\revtb{In evaluating the two use cases of VALERIAN, we present the performance of label correction only on the two controlled datasets with artificially added noise as their ground truth labels are available. For domain adaptation to an unseen subject, results are presented on all four datasets.}

\begin{figure}[h!]
	\centering
	\includegraphics[width=0.49\linewidth]{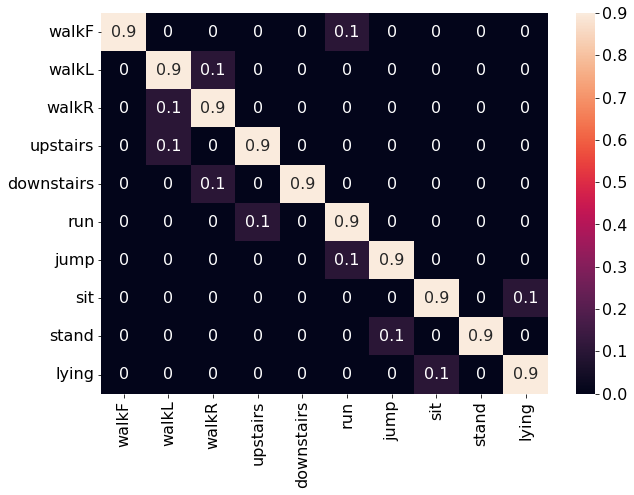}
	\includegraphics[width=0.49\linewidth]{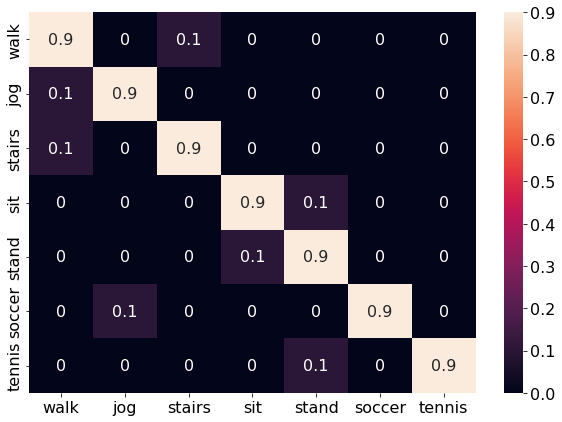}
	\caption{Noise transition matrix $T$ with asymmetric noise for USCHAD (Left) and WISDM (Right), $\tau=0.1$.}
	\label{fg:asym_t}
\end{figure}

Artificially injecting noise to clean labeled data is commonly used in evaluating LNL methods. For the controlled datasets, we consider
two noise patterns with four levels each, namely, symmetric noise with $\tau=\{0.1,0.2,0.4,0.6\}$ and asymmetric noise with $\tau=\{0.1,0.2,0.3,0.4\}$. The noise transition matrices \revtk{for} asymmetric ones are then defined according to Fig.~\ref{fg:asym_t}. From Section~\ref{sec:motivation}, we have seen that LNL with asymmetric noises is generally harder than that with symmetric noises. For example, when $\tau=0.4$ and the number of classes $C=10$ under asymmetric noise, roughly 60\% of data in each class is correctly labeled while the remaining 40\% is labeled to another class. As a result, the percentage difference between correctly and wrongly labeled data is only 20\%. In contrast, in the symmetric noise cases, the percentage gap is $60-\frac{40}{9} \approx 55.6\%$ (since the percentage of the wrongly labeled class is $\frac{40}{9}$). Therefore, for asymmetric noise, the maximum $\tau$ is set to 0.4 but in the case of symmetric noise, the maximum $\tau$ is set to to 0.6.
In the experiments, to better simulate real-world noise patterns, the noise transition matrices of asymmetric noise are defined by setting the probability of the most similar class of each activity to $\tau$, as shown in Fig.~\ref{fg:asym_t}\footnote{The most similar class is determined by the confusion matrix of a model trained on clean data.}. 

Leave-one-subject-out evaluation is conducted on all \revtb{four} datasets. In the experiments, we randomly select one subject as the target domain at a time, until all subjects are chosen. 
In Butterfly, we evaluate it in a way described in the original paper~\cite{liu2019butterfly} and \revtb{take 75\% of unlabeled target samples for training and the remaining for testing}. 
Experiments are repeated five times for each parameter setting, and the average test accuracy and its standard deviation are reported. 
\subsection{Results}
\label{sc:results}

\subsubsection{Label Correction for Source Domain} In this experiment, we evaluate the label correction accuracy and the overall accuracy of source domain data using the two controlled datasets with artificially injected symmetric and asymmetric noise. Here, we consider a wrongly label sample as a positive sample and thus recall is defined as the ratio between the number of correctly predicted samples that were previously wrongly labeled and the total number of wrongly labeled samples. 

\begin{figure}[h!]
	\centering
	\includegraphics[width=0.49\linewidth]{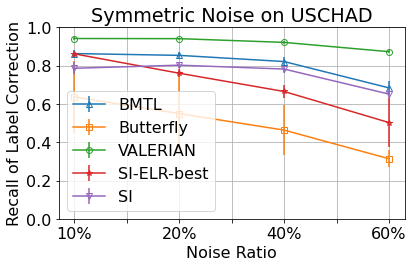}
	\includegraphics[width=0.49\linewidth]{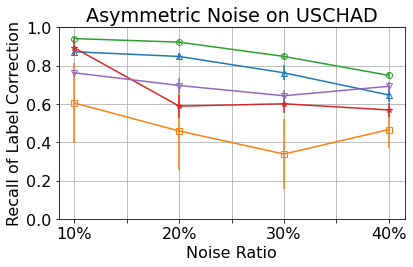}
	\includegraphics[width=0.49\linewidth]{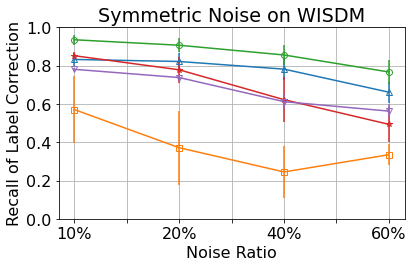}
	\includegraphics[width=0.49\linewidth]{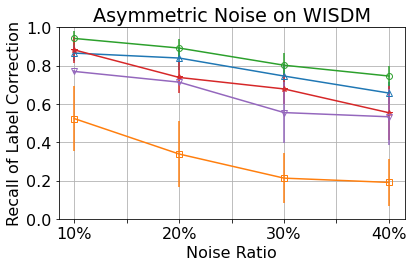}
	\caption{The recall rates of different methods on noisy training data.}
	\label{fg:clean_source_recall}
\end{figure}

\revtr{Fig. \ref{fg:clean_source_recall} shows the recall rates of different approaches.  It can be observed that VALERIAN outperforms all baseline methods by a large margin in all cases and can \revtk{correct} as high as 93\% of labelling errors when the noise level is 10\%. At high noise levels, e.g., 40\% asymmetric noise, its performance dropped to around 75\%. SI and SI-ELR have similar performance, both upper bounded by BMTL as they ignore the domain gap among training subjects. Interestingly, Butterfly performs the worst. This can be attributed to the fact that Butterfly treats data from different subjects as a single domain.}

\begin{figure}[h!]
	\centering
	\includegraphics[width=0.49\linewidth]{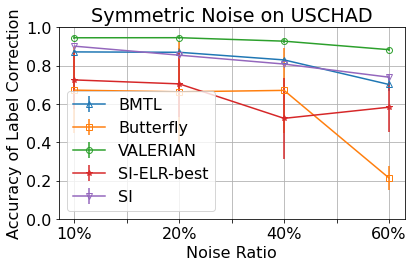}
	\includegraphics[width=0.49\linewidth]{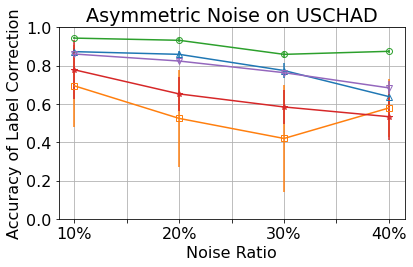}
	\includegraphics[width=0.49\linewidth]{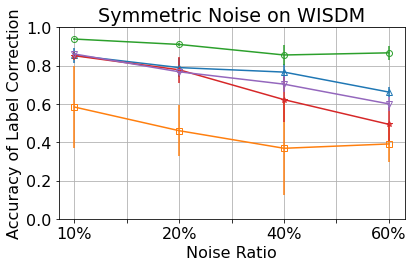}
	\includegraphics[width=0.49\linewidth]{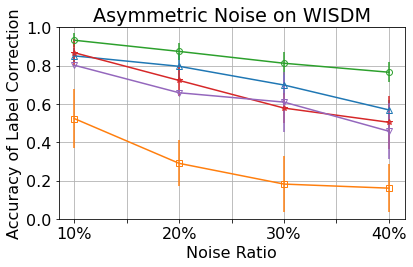}
	\caption{The training accuracy of different methods on noisy training data.}
	\label{fg:clean_source_accuracy}
\end{figure}

\revtr{Fig.~\ref{fg:clean_source_accuracy} shows the training accuracy for different methods. Benefiting from high recall rates of noisy data, VALERIAN achieves the best training accuracy among all. Note that the prediction errors in this setting include both mis-prediction of wrongly labeled data (i.e., memorization) and that of correctly labeled data in the training data, which can be due to the inherent limitation of the model architecture and uncorrected noisy labels. }
	
\subsubsection{Domain Adaptation with Clean Labeled Target Domain} 
First, we present the evaluation results on controlled datasets where clean data from unseen subjects is available. \revtb{As the case of symmetric noise is simpler, we only present results from asymmetric noise due to space limit.}

\begin{figure}[h!]
	\centering
	\includegraphics[width=0.49\linewidth]{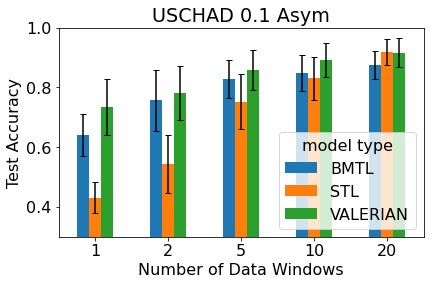}
	\includegraphics[width=0.49\linewidth]{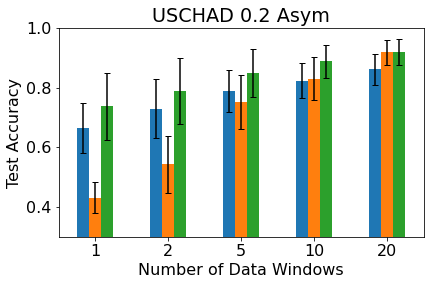}
	\includegraphics[width=0.49\linewidth]{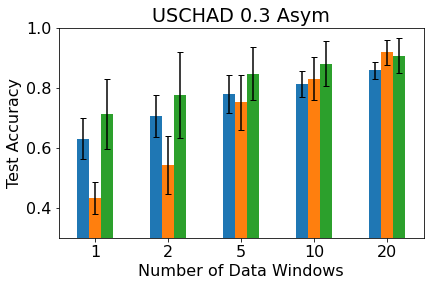}
	\includegraphics[width=0.49\linewidth]{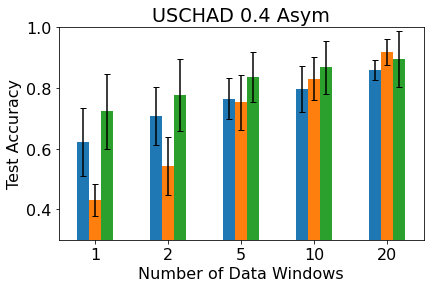}
	\caption{Evaluation on USCHAD with different levels of asymmetric noise and different numbers of data windows per activity class from $\mathcal{D}_t$. The test accuracy and standard deviation are averaged across all subjects in leave-one-out experiment.}
	\label{fg:uschad_asym}
\end{figure}

 
\begin{figure}[h!]
	\centering
	\includegraphics[width=0.49\linewidth]{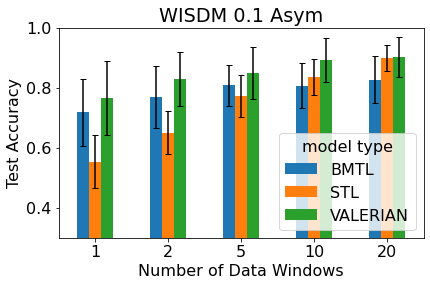}
	\includegraphics[width=0.49\linewidth]{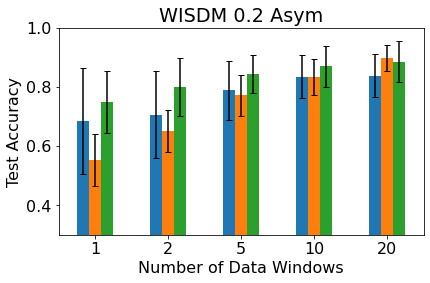}
	\includegraphics[width=0.49\linewidth]{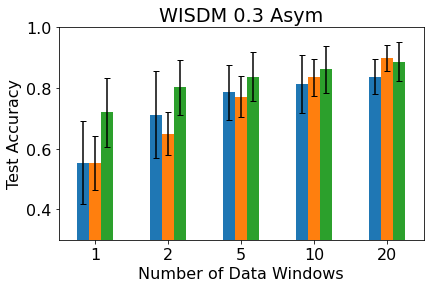}
	\includegraphics[width=0.49\linewidth]{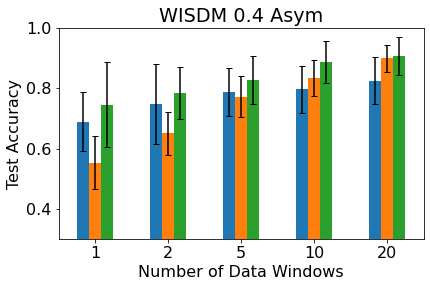}
	\caption{Evaluation on WISDM with different level of asymmetric noise  and different numbers of data windows per activity class from $\mathcal{D}_t$. The test accuracy and standard deviation are averaged across all subjects in leave-one-out experiment.}
	\label{fg:wisdm_asym}
\end{figure}

\paragraph*{Overall performance} 
Fig.~\ref{fg:uschad_asym} and ~\ref{fg:wisdm_asym} show the results on USCHAD and WISDM with asymmetric noise of different levels, respectively. From these figures, we observe that VALERIAN works well and its performance is quite stable across different noise levels and types of noise in both datasets. 
As STL is trained entirely on clean data from $\mathcal{D}_t$, its performance is not impacted by noise patterns and levels. As more clean data become available, the performance of STL serves as an upper bound of LNL models. From the figures, we see that with 20 shots, VALERIAN has comparable or slightly worse performance than STL. However, with a smaller number of target domain data, VALERIAN learns more efficiently. For example, with five shots, the average accuracy of VALERIAN for UHSCHAD and WISDM across all noise levels and patterns are 84.35 and 83.87, respectively, which are superior than BMTL (78.71 and 78.46) and STL (75.26 and 77.20). 
%
As the noise level increases, the accuracy of VALERIAN degrades slightly as expected. However, even with 40\% symmetric noise, it can achieve an average accuracy of 81.99\% for USCHAD for 5-shot learning, amounting to less than 3\% reduction compared to the case with 10\% symmetric noise. Similar observations can be made for asymmetric noise and WISDM. 


\paragraph*{Comparison with other LNL methods}
Table~\ref{tb:non_meta} summary the results of SI-ELR and Butterfly. For comparison, we also include the results of SI to further demonstrate that LNL methods can be harmful if applied naively. \revtb{SI, SI-ELR and VALERIAN are tested } with 5-shot learning \revtb{while Butterfly is a unsupervised domain adaptation method, which already sees  unlabeled target data during model training.} 

\begin{table}[t!]
	\caption{Results of methods \revtb{that are not designed for few-shot learning} on UHCHAD and WISDM, when 5 clean labeled samples per class are available from the target domain. Report with test accuracy in (\%).}
	\begin{subtable}{\linewidth}
		\resizebox{\linewidth}{!}{%
			\begin{tabular}{l|llll}
				
				\hline
				Method/Noise ratio & 10\% & 20\% & 30\% & 40\% \\ \hline
				SI         & $74.26 \pm 16.07$      & $73.44 \pm 13.57$     & $68.66 \pm 13.56$     & $65.35 \pm 14.31$        \\ \hline
			
				SI-ELR-best               & $77.84 \pm 15.41$    &   $70.48 \pm 17.38$    &$69.71 \pm 20.43 $      &   $58.49 \pm 13.17$        \\ \hline
				Butterfly             & $65.14 \pm 15.25$     & $52.44 \pm 25.51$     & $41.96 \pm 28.15$     &   $37.89 \pm 15.06$       \\ \hline \hline
				VALERIAN           & $85.71 \pm 6.65$     & $84.88 \pm 8.11$     & $84.81 \pm 8.82$     & $83.68 \pm 8.18$            \\ \hline
			\end{tabular}
		}
		\caption{Results on USCHAD dataset with four levels of artificially added \revtr{asymmetric noise} patterns in data labels.}
	\end{subtable}

	\begin{subtable}{\linewidth}
		\resizebox{\linewidth}{!}{%
			\begin{tabular}{l|llll}
				
				\hline
				Method/Noise ratio  & 10\% & 20\% & 30\% & 40\% \\ \hline
				SI       & $58.66 \pm 17.36$      & $56.81 \pm15.29$     & $53.10 \pm 13.73$   & $48.38 \pm 11.36$        \\ \hline
				SI-ELR-best                & $66.15 \pm 11.46$     & $65.23 \pm 8.85$     &$58.59 \pm 11.91$      & $55.10 \pm 9.66$          \\ \hline
				Butterfly              & $57.98 \pm 14.66$     & $36.75 \pm 13.23$     & $24.47 \pm 15.36$     & $14.30 \pm 1.43$            \\ \hline \hline
				VALERIAN             & $84.85 \pm 8.73$     & $84.41 \pm 6.53$     & $83.71 \pm 8.01$     & $82.63 \pm 7.98$            \\ \hline
			\end{tabular}
		}
		\caption{Results on WISDM dataset with four levels of artificially added \revtr{asymmetric noise} patterns in data labels.}
	\end{subtable}

	\label{tb:non_meta}
\end{table}
From Table \ref{tb:non_meta}, it is clear that none of the three methods performs well in HAR with noisy labels. The vanilla SI model does not explicitly handle subject divergence nor label noises. Its performance degrades as the noise ratio $\tau$ increases. In comparison, SI-ELR ignores subject divergence and deals with noisy labels using a regularization term. 
Though designed to handle label noise, SI-ELR-best has worse performance than SI when the asymmetric noise level is greater than 10\%. The results are consistent with our observations with DivideMix in Section~\ref{sec:motivation} and reveal that subject diversity harms ELR's ability to combat label noises. SI-ELR fares moderately better for asymmetric noise. However, with 40\% noise, SI-ELR-best is 7\% worse than SI and 25\% worse than VALERIAN in USCHAD. 

Butterfly on average has worse accuracy than SI and SI-ELR-best and performs poorly as the noise level increases in both datasets. This is in part due to the fact that Butterfly uses unlabeled target domain data at training time while SI and SI-ELR-best benefit from transfer learning with a few shots of clean labeled data at inference time. However, the difference in accessing target domain labels does not justify the large variance in Bufferfly's test accuracy on USCHAD as shown in Table~\ref{tb:non_meta}. \revtr{As an example, with 0.3 asymmetric noise, its highest test accuracy is 70.11\% when subject 7 is in the test set, whereas its lowest accuracy is 13.8\% for test subject 8. }
We believe that the poor performance of Butterfly is because it treats different subjects in the training set as a single domain. 

\paragraph*{Ablation Study} To see how each component contributes to the final performance of VALERIAN, an ablation study \revtk{was} conducted on the USCHAD dataset with 5-shot learning and 0.4 asymmetric noise. Similar results \revtk{could} be expected for other noise settings or datasets.
\begin{table}[t!]
\caption{Ablation study of VALERIAN on USCHAD with 0.4 asymmetric noise.}
\begin{tabular}{l|l}
		\hline
		{\bf Method}                               & {\bf Test Accuracy} \\ \hline
		VALERIAN                               &   $83.68 \pm 8.18$            \\ \hline
		VALERIAN w/o ELR                       &   $76.55 \pm 6.69$            \\ \hline
		VALERIAN w/o self-supervised pre-train &   $79.28 \pm 7.37$            \\ \hline
		VALERIAN w/o MixUp                     &   $77.69 \pm 2.34$            \\ \hline
		VALERIAN w/o IFLF                      &  $60.18 \pm 10.35$             \\ \hline
	\end{tabular}

\label{tb:ablation}
\end{table}
As shown in Table. \ref{tb:ablation}, the domain invariant feature learner plays the most important role in VALERIAN. Without IFLF, VALERIAN degrades to an ELR model and fails to deal with subject divergence. Moreover, in absence of a dedicated meta-learning strategy,  it is insufficient to update parameters of the whole model by only a few clean labeled data samples. As a result, a large standard deviation in test accuracy is observed. MixUp contributes a $\sim6\%$ accuracy to the overall solution, empirically demonstrating its usefulness in improving model generalization in HAR tasks with noisy labels. Inclusion of ELR in VALERIAN leads to $\sim7\%$ improvement. Recall the poor performance of ELR alone in Table~\ref{tb:non_meta}. The results speak unequivocally for the need to combine LNL and meta-learning to handle subject diversity. Lastly, we find that self-supervised pre-train contributes $\sim4\%$ test accuracy.

\subsubsection{Domain Adaptation on Noisy Labeled Target} 
Next, we compare the performance of VALERIAN, BMTL and STL on two noisy labeled datasets: ExtraSensory \revtb{and RealWorld}, which 
\revtb{are in the wild datasets.} Considering the data imbalance and class missing issue, we take F1-Score rather than accuracy as metrics to evaluate model performance here. Note that since the ground truth labels from curated data are noisy, the quantitative results need to be taken with a grain of salt. \revtb{To generate t-SNE plots, we randomly selected one subject from each dataset and cleaned its labels manually.}

\begin{figure*}[h!]
	\centering
		\begin{subfigure}{0.32\linewidth}
		\includegraphics[width=\textwidth]{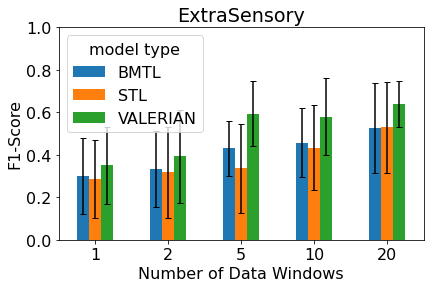}
		\caption{F1-Score.}
		\end{subfigure}
		\begin{subfigure}{0.32\linewidth}
		\includegraphics[width=\textwidth]{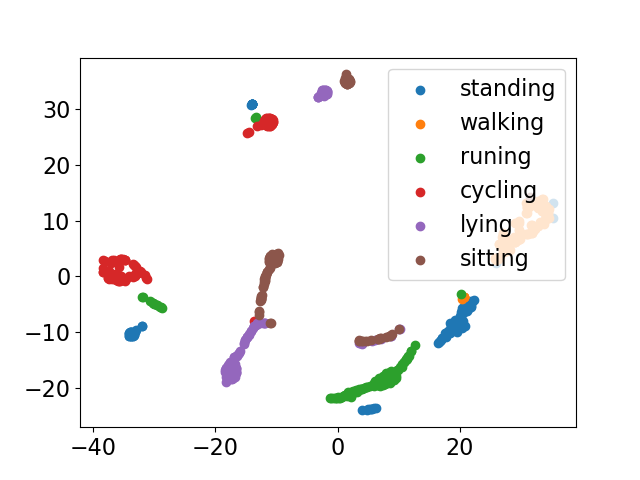}
		\caption{t-SNE plot of features in BMTL.}
		\end{subfigure}
		\begin{subfigure}{0.32\linewidth}
		\includegraphics[width=\textwidth]{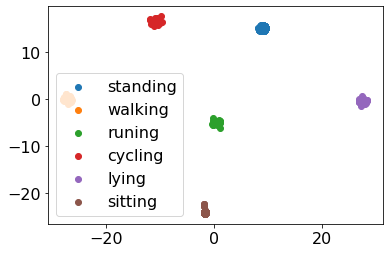}
		\caption{t-SNE plot of features in VALERIAN.}
		\end{subfigure}
	\caption{Evaluation on ExtraSensory with different the number of data windows per activity class from $\mathcal{D}_t$. The mean and standard deviation F1-Scores are averages across all subjects in leave-one-out experiment. t-SNE are generated on a random subject (id:4FC32141-E888-4BFF-8804-12559A491D8C) with data from all six classes.}
	\label{fg:wild_result}
\end{figure*}

Fig. \ref{fg:wild_result}(a) shows the F1-Score of the three models with gradually increasing the number of data windows on ExraSensory . 
Compared to results with the two controlled datasets, all methods show their worst performance. This can be attributed to the noisy target domain labels \revtk{during} fast adaption or learning STL model. The large standard deviation in STL results even \revtk{with} 20-shots indicates either label noise in target domain data or noise in ground truth or both. In fact, for many \revtk{of the} subjects, the training and validation set is not i.i.d due to data noise, resulting in a validation accuracy jumping back and forth between training epochs. However, VALERIAN still outperforms the other two methods in all cases \revtk{tested}. 
To see if VALERIAN can indeed learn good features from noisy data, we show in Fig. \ref{fg:wild_result}(c) the t-distributed stochastic neighbor embedding (t-SNE) plot of the outputs of its feature extraction network. Clearly, the classes are well separated. This is in contrast with overlapping among classes in \ref{fg:wild_result}(b), which shows the t-SNE plot of the outputs from the feature extraction network in BMTL. 

\begin{figure*}[h!]
	\centering
		\begin{subfigure}{0.33\linewidth}
		\includegraphics[width=\textwidth]{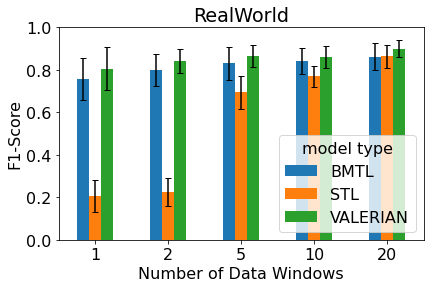}
		\caption{F1-Score.}
		\end{subfigure}
		\begin{subfigure}{0.33\linewidth}
		\includegraphics[width=\textwidth]{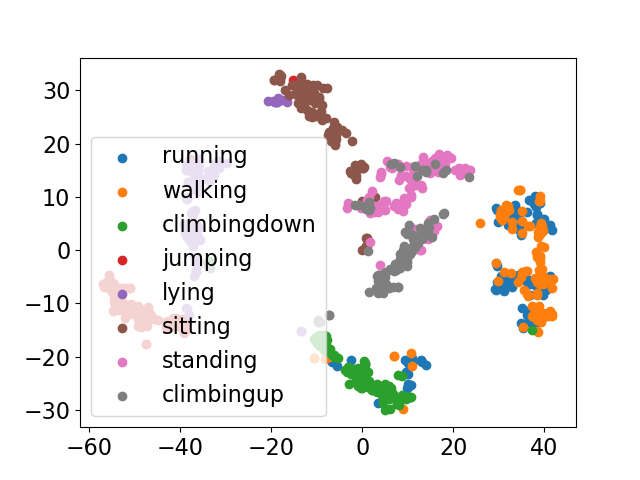}
		\caption{t-SNE plot of features in BMTL.}
		\end{subfigure}
		\begin{subfigure}{0.33\linewidth}
		\includegraphics[width=\textwidth]{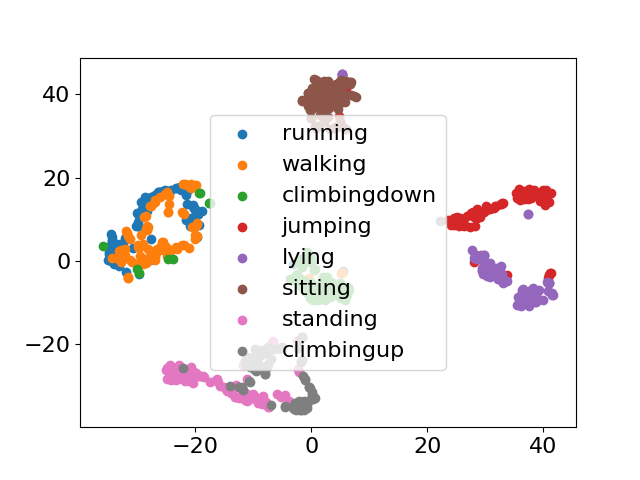}
		\caption{t-SNE plot of features in VALERIAN.}
		\end{subfigure}
	\caption{Evaluation on RealWorld with different the number of data windows per activity class from $\mathcal{D}_t$. The mean and standard deviation F1-Scores are averages across all subjects in leave-one-out experiment. t-SNE are generated on a random subject (id:3) with data from all eight classes.}
	\label{fg:realworld_result}
\end{figure*}
\revtb{Similar observations can be made for RealWorld. With BMTL, the t-SNE plot in Figure~\ref{fg:realworld_result}(b) shows closeness (and thus likely mislabeling) between running and walking, standing and climbing up, climbing down and walking activities. In comparison, clusters generated by VALERIAN are better separated.}

\section{Related Work}
\label{sc:RelatedWork}
\paragraph*{Learning with noisy labels} LNL has been investigated in computer vision and audio signal processing for over a decade \cite{han2020survey,song2022learning}. Existing methods can be categorized into three groups. First, contrastive learning-based LNL methods \cite{yi2022learning,yang2021partially} add regularization terms to the loss function to obtain a well-clustered feature structure. Second, curriculum learning \cite{braun2017curriculum,lyu2019curriculum} or teacher-student networks such as MentorNet \cite{jiang2018mentornet} trains a neural network to guide a student network by assigning weights to samples. Since the pioneer co-teaching work~\cite{han2018coteach}, the use of two networks together gains popularity in  LNL and has been adopted in several recent papers including DivideMix \cite{li2019dividemix}, ELR+ \cite{liu2020early}, co-regularization \cite{wei2020combating}). Instead of training a model that works on the noisy labeled samples, another line of work aims to select clean labeled samples out of noisy ones \cite{northcutt2021confident,zhang2020distilling}. Despite all the advancements in  LNL, none of the afore-mentioned work considers domain gaps between source and target domains (also known as {domain shifts}).
\paragraph*{Weakly-supervised learning in sensor-based HAR} There are some works in  mobile computing  that deal with weakly-supervised learning problems related to sensor-based HAR \cite{he2018weakly,wang2021sequential,wang2019attention}. Wang {\it et al.} in \cite{wang2021sequential,wang2019attention} define weakly-supervised learning as detecting the start and end of an activity of interest in a given time-series data sequence, similar to the sound event detection problem\cite{kumar2016audio,dinkel2021towards,adavanne2019sound}. Unlike our problem, the goal is to crop the data of interest from a noisy sequence for training so that a machine learning model can gain a better discriminative power. For instance, consider a collected \textit{climbing up} IMU data trial with two activities: climbing upstairs and walking on the flat ground. Wang {\it et al.} treat walking as a background  activity and try to detect the onset and offset timestamps of climbing upstairs events. In contrast, in this work, we treat the data within such a trial as a mixture of \textit{climbing up} and noisy labeled \textit{walking} activities. Apart from the different ways of treating label noises, existing works still require further steps to handle subject diversity within the training process to generalize well to new unseen subjects.
\paragraph*{Joint LNL and domain adaptation} A few works consider LNL together with domain shifts. Shu {\it et al.} in \cite{shu2019transferable} considered noise either in data or label of a single source domain and perform weakly-supervised model training to adapt to a target domain. In \cite{liu2019butterfly,yu2020label} researchers propose one-step solutions to LNL and unsupervised domain adaptation. However, these methods have been applied to image classification tasks, where there is only a single source domain. Thus, the authors only consider the domain shift  between one source domain and one target domain. In contrast, in our work, we need to take into account domain shifts amongst multiple source domains, namely, different human subjects. As discussed in Section~\ref{sec:motivation}, subject diversity in training data prevents conventional LNL methods from working effectively since early learning can inadvertently memorize noisy data.  

\section{Conclusion}
\label{sc:Conclusion}
In this paper, we proposed VALERIAN, a domain invariant feature learning approach for \revtk{IMU }sensor-based HAR in the wild. An extensive experimental study demonstrated its superior performance over baseline methods for different levels of noise and noise patterns, and  \revtb{in two use scenarios}. The key takeaway from this work is two-fold: 1) the effects of subject diversity and label noises intertwine in the learning behaviour of LNL models and can lead to catastrophic memorization of wrongly labelled data, and 2) it is important to design domain adaptation strategies to explicitly handle subject diversity in conjunction with LNL for better generalization in HAR. 

\revtk{It is plausible to apply VALERIAN to other sensor data modalities as long as there exist significant subject divergence and performance drop due to label noises.}
Components of VALERIAN (e.g, self-supervised learning, early loss regularization) can be replaced by other more advanced approaches though the framework remains applicable. Also orthogonal to the proposed approach are unsupervised domain adaption methods and domain generalization methods. \revtk{One limitation of VALERIAN  is the need of correctly labeled samples from a target domain for domain adaptation to achieve a higher inference accuracy. One interesting area of further investigation is to perform domain adaptation with noisy labeled target only.} Finally, we believe significant efforts should be made to build in-the-wild datasets and benchmarks for IMU-based HAR. 


\bibliographystyle{ACM-Reference-Format}
\bibliography{ref}

\end{document}